\documentclass[aps,prb,superscriptaddress,reprint,twocolumn,notitlepage,showpacs]{revtex4}

\usepackage[T2A]{fontenc}
\usepackage[cp1251]{inputenc}
\usepackage{graphicx}
\usepackage{epsfig}
\usepackage{amsmath}
\usepackage{caption2}
\usepackage{floatflt}
\usepackage{dsfont}
\usepackage{mathrsfs}

\newcommand*{\be}{\begin{equation}}
\newcommand*{\ee}{\end{equation}}

\newcommand*{\IDcp}{\int D\chi\: D\bar\chi \:D\psi\: D\bar\psi\:}
\newcommand*{\Idtdx}{\int_0^{\hbar\beta}d\tau\int d{\bf x}}
\newcommand*{\x}{{\bf x}}
\newcommand*{\ra}{{\bf r}}
\newcommand*{\sa}{{\bf s}}
\newcommand*{\dx}{d{\bf x}}
\newcommand*{\p}{{\bf p}}

\begin{document}
\title{Coupled exciton-photon Bose condensate in path integral formalism}
\author{\firstname{A.~A.}~\surname{Elistratov}}
\email{elistr@rambler.ru}
\affiliation{%
 Institute for Nanotechnology in Microelectronics RAS, 119334 Leninskiy ave. 32a, Moscow, Russia
}
\author{\firstname{Yu.~E.}~\surname{Lozovik}}
\affiliation{%
 Institute for Spectroscopy RAS, 142190 Troitsk, Moscow, Russia
}

\begin{abstract}
We study the behavior of exciton polaritons in an optical microcavity with an embedded semiconductor quantum well. We use two-component exciton-photon approach formulated in terms of path integral formalism. In order to describe spatial distributions of the exciton and photon condensate densities, the two coupled equations of the Gross-Pitaevskii type  are derived. For a homogeneous system, we find the noncondensate photon and exciton spectra, calculate the coefficients of transformation from the exciton-photon basis to the lower-upper polariton basis, and obtain the exciton and photon occupation numbers of the lower and upper polariton branches  for nonzero temperatures. For an inhomogeneous system, the set of coupled equations of the Bogoliubov-de-Gennes type is derived. The equations govern the spectra and spatial distributions of noncondensate photons and excitons.   
\end{abstract}

\maketitle

\section{Introduction}
In the last two decades there has been extensive work on  
Bose-Einstein condensation in the system of two-dimensional exciton polaritons in optical microcavites (see the reviews \cite{Yamamoto,Pitaevskii,Kavokin1,Carusotto1} and the references therein).  The profound investigation of properties of the polariton system has taken place \cite{Ciuti1, Tassone, Eastham, Porras,Ciuti2,Keeling1,Kavokin2,Laussy,Kammann}. Bose condensation has been demonstrated \cite{Dang,Richard,Kasprzak,Balili,Lagoudakis1,Amo1,Amo2,Amo3} and many features of the polariton Bose condensate such as superfluidity  \cite{Carusotto2,Wouters1}, vortices \cite{Marchetti,Keeling2,Borgh1,Borgh2,Rubo,Hamp,Lagoudakis2,Krizhanovskii,Sanvitto,Tosi,Anton1,Voron2} and solitons \cite{Amo4,El,Flayac,Solnyshkov} formation were studied, and have been shown to have specific character in the non-equilibrium spatially inhomogenious system of exciton polaritons. Recently, novel phenomena such as ballistic transport over macroscopic distances \cite{Leyder,Wertz1}, spontaneous oscillations  \cite{Demirchyan,Lieb,Giorgi},  Josephson-like effects  \cite{Wouters2,Shelykh1,Sarchi,Read,Lagoudakis3,Abbarchi,Pavlovic},  optical Aharonov-Bohm \cite{Shelykh2} and spin Hall effects \cite{Kavokin3} are actively discussed.

Most authors of theoretical works use, explicitly or not, the conventional approach consisting of the following steps: one writes out the Hamiltonian of the system including exciton and photon creation and annihilation operators, performs the Hopfield canonical transformation from the exciton-photon to  the lower-upper polariton basis, neglects the upper polariton branch as empty at low temperatures, and regards polaritons on the lower polariton branch as a dilute Bose gas weakly interacting via the exciton component. Such an approach allowing the straightforward application of the Bogoliubov theory has led to a great amount of important results (see the broad review of the approach  in Ref. \cite{Carusotto1}).         

However, this approach has two serious shortages. The first one is the neglection of the upper polariton branch. It is quite justified at the temperature range of a few degrees Kelvin which takes place in up-to-date experiments. Though, as many authors have noted, Bose condensation of polaritons is so attractive due to extremely high transition temperature being the effect of very small polariton effective mass. Already for nitrogen temperatures, the upper polariton branch occupation is not fully negligible. Second and more serious shortage of the conventional approach is that the canonical transformation from the photon-exciton to the polariton basis is performed in the momentum space and hence it requires the spatial uniformity or quasi-uniformity of the system.       
 
With the up-to-date traps  having dimensions of the order of dozen microns \cite{Willander,Balili1,Balili2,Berman,Lozovik,Anton2}, one can perform the canonical transformation  at each point in space in the frame of the quasi-classical approximation. However, latest experiments have studied systems already with micron dimensions \cite{Wertz2,Anton3,Sigurdsson} and it is likely that the trend of miniaturization will go further. Besides, the Bose condensed system can exhibit topological defects such as vortices and solitons \cite{Marchetti,Keeling2,Borgh1,Borgh2,Rubo,Hamp,Lagoudakis2,Krizhanovskii,Sanvitto,Tosi,Anton1,Amo4,El,Flayac,Solnyshkov,Voron2}. The exciton and photon healing lengths, which determine the spatial scales of these defects, are of the order of micron. Thus, starting from the micron scale the conventional approach becomes non-applicable. Moreover, polariton as a well defined quasi-particle  at this scale  does not make sense.

In this paper, we study the behavior of exciton polaritons in the framework of a two-component exciton-photon approach. The exciton and photon components are described as independent quantum fields interacting via Rabi splitting. Bose condensation leads to the  appearance of the two coupled exciton and photon condensates. In the mean field approximation this approach has been used, for example,  in Refs.\cite{Whittaker,Hamp,Voron1,Voron2,Voron3}. The authors of Refs.\cite{Carusotto2,Ciuti3} have gone beyond the mean field approximation and calculated the energy spectrum of noncondensate photons and excitons. 

We analyze the equilibrium Bose condensed polariton system without pumping and photon decay from the cavity at nonzero, but sufficiently small temperature. These effects were studied in the two-component approach in Ref. \cite{Ciuti3}. Moreover, we do not take into account the spin degree of freedom of polaritons.

The approach is formulated in terms of path integral formalism for the following reasons. First, this formalism allows the study of  both condensate and noncondensate parts of the system from the unified point of view. The second advantage is the use of coherent states, which are the natural choice for an exploration of quantum properties of light emitted from the polariton system. After all, the method involves Matsubara technique and hence it is proved to be effective for a study of the polariton behavior at nonzero temperatures. The application of the path integral method to a weakly interacting condensed Bose gas can be found in the excellent review by Stoof \cite{Stoof1}. 

The paper is organized as follows. In Section II we describe the geometry of the polariton system and introduce the exciton and photon one-particle eigenstates. In Section III we write the partition function of the system as a path integral, which contains an action fully describing our system. The exciton and photon quantum fields in the path integral formalism are c-number fields written in the basis of coherent states. In Section IV we turn to the Bose condensed polariton system and write out both exciton and photon fields as a sum of the order parameter and fluctuations around it. The part of the action, which does not contain fluctuations, is the Pitaevskii-like functional. Its minimization leads to the two coupled equations of the Gross-Pitaevskii type. These equations give the description of the system of the two coupled condensates in the mean field approximation. We analyze the properties of the obtained equations and find the conditions of equivalence between the one-component and two-component approaches. In order to explore the noncondensate part of the system, in Section V we analyze the quadratic in fluctuations part of the action. The most important result here is the canonical transformation from the exciton-photon basis to the lower-upper polariton basis. This transformation incorporates both the Hopfield polariton transformation and the Bogoliubov transformation for weakly-interacting Bose gas. Moreover, we obtain noncondensate particles energy spectra and calculate exciton and photon occupation numbers for lower and upper polariton branches at nonzero temperatures. In Section VI we derive the set of coupled equations of the Bogoliubov-de-Gennes type. This equations govern the energy spectra and spatial distributions of the noncondensate excitons and photons in a spatially inhomogeneous system.

\section{Geometry and one-particle eigenstates}
We study a system of exciton polaritons in a semiconductor optical microcavity with an embedded quantum well. In the simple case when the quantum well possesses in-plane translational symmetry, energy spectrum of excitons in the region of small in-plane momenta has the form
\be
\varepsilon^{ex}=\varepsilon_{(0)}^{ex}+\frac{\p^2}{2m_{ex}},
\ee 
where $\varepsilon$ is the dielectric constant of the medium, $\varepsilon_{(0)}^{ex}=2m_{ex}e^4/\varepsilon^2\hbar^2$ is the 2D exciton binding energy, $m_{ex}=m_e+m_h$ is the 2D exciton mass, and $m_e$ and $m_h$ are the effective masses of an electron and a hole, respectively.   
 
An external confining potential $V^{ex}(\x)$ for excitons in the quantum well can be created. Experimentally, there are several methods to realize the  confinement of excitons. One of them is the exciton energy shifting using a stress-induced band-gap shift \cite{Willander, Balili, Berman}.    
 
We assume one-particle eigenstates describing noninteracting excitons confined by the external potential $V^{ex}(\x)$ to be found from the time-independent Schr$\ddot o$dinger equation 
\be
\left(-\frac{\hbar^2\bigtriangledown^2}{2m_{ex}}+V^{ex}(\x)-\varepsilon_{\bf n}^{ex}\right)\chi_{\bf n}(\x)=0.
\ee
The set of solutions ${\chi_{\bf n}(\x)}$ is orthonormalized 
\be
\int \dx \:\bar\chi_{\bf n}(\x)\chi_{\bf m}(\x)=\delta_{{\bf nm}}
\ee
and full
\be\label{sumchin}
\sum_{\bf n} \chi_{\bf n}(\x)\bar\chi_{\bf n}(\x')=\delta(\x-\x').
\ee
In the case of the planar microcavity the photons in the region of small in-plane momenta have the following dispersion:  
\be\label{eph}
\varepsilon^{ph}=\frac{\hbar c}{\sqrt{\varepsilon}}\sqrt{k_\perp^2+k^2}\approx \frac{\pi\hbar c}{L\sqrt{\varepsilon}}n+\frac{\p^2}{2m_{ph}}.
\ee
Here $m_{ph}=\pi\hbar\sqrt{\varepsilon}/cL$ is the effective photon mass. We consider the lowest state $n=1$. 
   
There are several experimental approaches to realize the  confinement of photons \cite{Lozovik}. It is, for example, creating a special dielectric permittivity profile inside the cavity which would lead to photon localization, or  making a trap for photons based on special shaping of the microcavity width $L(\x)$ (up to creation of a finite system limited by mirrors from all sides). In this case the effective photon mass $m_{ph}$ and the first term in (\ref{eph}) become functions of coordinates. As a result, the following one-particle problem arises for photons in the microcavity with a nonconstant width $L(\x)$: 
\begin{multline}
-\frac{\hbar^2}{2}\nabla\left(\frac{1}{m_{ph}(\x)}\nabla\psi_{\bf n}(\x)\right)\\
+\left(\frac{\pi\hbar c}{L(\x)\sqrt{\varepsilon}}-\varepsilon_{\bf n}^{ph}\right)\psi_{\bf n}(\x)=0.
\end {multline}
We will assume microcavities to possess in-plane translational symmetry, and will not consider microcavities with photon confinement, i.e. $m_{ph}(\x)=m_{ph}=const$.  The one-particle problem for photons reduces to
\be
\left(-\frac{\hbar^2\bigtriangledown^2}{2m_{ph}}-\varepsilon_{\bf n}^{ph}\right)\psi_{\bf n}(\x)=0.
\ee
Similarly to the above exciton one-particle eigenstates, we assume the set of solutions of this problem to be orthonormalized and full
\be
\int \dx \:\bar\psi_{\bf n}(\x)\psi_{\bf m}(\x)=\delta_{{\bf nm}},
\ee
\be
\sum_{\bf n} \psi_{\bf n}(\x)\bar\psi_{\bf n}(\x')=\delta(\x-\x').
\ee

\section{Action and free-field Green's functions}
In the functional approach, the grand-canonical function of the exciton-photon system is the functional integral 
\be\label{Z}
Z=\IDcp e^{-\frac{1}{\hbar}S[\bar\chi,\chi,\bar\psi,\psi]},
\ee
where the action $S$ consists of four terms 
\begin{multline}\label{S}
S[\bar\chi,\chi,\bar\psi,\psi]=S_{ex}[\bar\chi,\chi]+S_{ph}[\bar\psi,\psi]\\
+S_{RABI}[\bar\chi,\chi,\bar\psi,\psi]+S_{int}[\bar\chi,\chi],
\end{multline}
describing excitons, photons, Rabi splitting and interexciton interaction, respectively:
\begin{multline}
S_{ex}[\bar\chi,\chi]=\Idtdx\:\:\bar\chi(\x,\tau)\\ \times\left(\hbar\frac{\partial}{\partial\tau}-\frac{\hbar^2\bigtriangledown^2}{2m_{ex}}
+V^{ex}(\x)+E_0-\mu\right)\chi(\x,\tau),
\end{multline}
\begin{multline}
S_{ph}[\bar\psi,\psi]=\Idtdx\:\:\bar\psi(\x,\tau)\\ \times\left(\hbar\frac{\partial}{\partial\tau}-\frac{\hbar^2\bigtriangledown^2}{2m_{ph}}
-\mu\right)\psi(\x,\tau),
\end{multline}
\begin{multline}
S_{RABI}[\bar\chi,\chi,\bar\psi,\psi]=\Idtdx\\
\times\frac{\hbar\Omega}{2}\left[\bar\psi(\x,\tau)\chi(\x,\tau)
+\psi(\x,\tau)\bar\chi(\x,\tau)\right],
\end{multline}
\begin{multline}
S_{int}[\bar\chi,\chi]=-\frac{1}{2\hbar}\Idtdx\:d{\bf x'}\:\bar\chi(\x,\tau)\bar\chi(\x',\tau)\\
\times V(\x-\x')\chi(\x',\tau)\chi(\x,\tau).
\end{multline}
Here $\chi(\x,\tau)$ and $\psi(\x,\tau)$ are the field operators written in the basis of coherent states and hence they constitute c-number functions connected with $\chi_n(\x)$ и $\psi_n(\x)$ via relations
\be\label{chipsi}
\chi(\x,\tau)=\sum_{n}\chi_n(\tau)\:\chi_n(\x),\quad
\psi(\x,\tau)=\sum_{n}\psi_n(\tau)\:\psi_n(\x) 
\ee 
with the coefficients of expansion $\chi_n(\tau)$ and $\psi_n(\tau)$ dependent on time. 

We include the detuning $E_0$ between the exciton spectrum $\varepsilon_{\p}^{ex}=E_0+{\p}^2/2m_{ex}$ and photon one $\varepsilon_{\p}^{ph}={\p}^2/2m_{ph}$ into $S_{ex}$ and consider $E_0>0$.

Further we will assume the exciton interaction to have the approximate form $V(\x-\x')=V_0\:\delta(\x-\x')$. 

The partition function $S_{ex}[\bar\chi,\chi]$ can be rewritten as
\begin{multline}
-\frac{S_{ex}[\bar\chi,\chi]}{\hbar}=\Idtdx\int_0^{\hbar\beta}d\tau'\int \dx'\\
\bar\chi(\x,\tau)G_{ex(0)}^{-1}(\x,\tau;\x',\tau')\chi(\x',\tau'),
\end{multline}
where
\begin{multline}\label{Gex0-1}
G_{ex(0)}^{-1}(\x,\tau;\x ',\tau')
=-\frac{1}{\hbar}\left(\hbar\frac{\partial}{\partial\tau}-\frac{\hbar^2\bigtriangledown^2}{2m_{ex}}\right.\\
\left.+V^{ex}(\x)+E_0-\mu\biggr)\delta(\x-\x')\delta(\tau-\tau')\right. 
\end{multline}
or, equivalently,
\begin{multline}\label{Gex0}
\left(\hbar\frac{\partial}{\partial\tau}-\frac{\hbar^2\bigtriangledown^2}{2m_{ex}}+V^{ex}(\x)+E_0-\mu\right)\\ \times G_{ex(0)}(\x,\tau;\x ',\tau')
=-\hbar\delta(\x-\x')\delta(\tau-\tau').
\end{multline}
Substituting $\delta(\x-\x')$ from (\ref{sumchin}) and using the equality 
\be
\delta(\tau-\tau')=\frac{1}{\hbar\beta}\sum_m e^{-i\omega_m (\tau-\tau')}
\ee 
we obtain
\begin{multline}
G_{ex(0)}^{-1}(\x,\tau;\x ',\tau')=
-\frac{1}{\hbar}\sum_{n,m}(-i\hbar\omega_m + \varepsilon_n^{ex}+E_0-\mu)\\ \times\chi_n(\x)\bar\chi_n(\x')\frac{e^{-i\omega_m (\tau-\tau')}}{\hbar\beta}.
\end{multline}
The solution of the equation (\ref{Gex0}) reads
\begin{multline}
G_{ex(0)}(\x,\tau;\x ',\tau')=\sum_{n,m}\frac{-\hbar}{-i\hbar\omega_m + \varepsilon_n^{ex}+E_0-\mu}\\
\times\chi_n(\x)\bar\chi_n(\x')\frac{e^{-i\omega_m (\tau-\tau')}}{\hbar\beta}
\end{multline}
Since the functions $G_{ex(0)}^{-1}$ and $G_{ex(0)}$ are mutually inverse, they obey the relation $G_{ex(0)}G_{ex(0)}^{-1}=I$, where $I$ is the unity matrix. In coordinate space, this relation has the form 
\begin{multline}
\int_0^{\hbar\beta}d\tau''\int d{\bf x''}\:G_{ex(0)}(\x,\tau;\x'' ,\tau'')\\
\times G_{ex(0)}^{-1}(\x'',\tau'';\x ',\tau')
=\delta(\x-\x')\delta(\tau-\tau').
\end{multline}
Similarly, $S_{ph}[\bar\psi,\psi]$ can be written as
\begin{multline}
-\frac{S_{ph}[\bar\psi,\psi]}{\hbar}=\Idtdx\int_0^{\hbar\beta}d\tau'\int \dx '\\
\times\bar\psi(\x,\tau)G_{ph(0)}^{-1}(\x,\tau;\x',\tau')\psi(\x',\tau'),
\end{multline}
where
\begin{multline}\label{Gph0-1}
G_{ph(0)}^{-1}(\x,\tau;\x ',\tau')=\\
-\frac{1}{\hbar}\left(\hbar\frac{\partial}{\partial\tau}-\frac{\hbar^2\bigtriangledown^2}{2m_{ph}}-\mu\right)\delta(\x-\x')\delta(\tau-\tau').
\end{multline}
This equation is equivalent to
\begin{multline}\label{Gph0}
\left(\hbar\frac{\partial}{\partial\tau}-\frac{\hbar^2\bigtriangledown^2}{2m_{ph}}-\mu\right)G_{ph(0)}(\x,\tau;\x ',\tau')=\\
-\hbar\delta(\x-\x')\delta(\tau-\tau').
\end{multline}
For the functions $G_{ph(0)}(\x,\tau;\x ',\tau')$ and $G_{ph(0)}^{-1}(\x,\tau;\x ',\tau')$ we have
\begin{multline}
G_{ph(0)}^{-1}(\x,\tau;\x ',\tau')=-\frac{1}{\hbar}\sum_{n,m}(-i\hbar\omega_m + \varepsilon_n^{ph}-\mu)\\
\times\psi_n(\x)\bar\psi_n(\x')\frac{e^{-i\omega_m (\tau-\tau')}}{\hbar\beta},
\end{multline}
\begin{multline}
G_{ph(0)}(\x,\tau;\x ',\tau')=\sum_{n,m}\frac{-\hbar}{-i\hbar\omega_m + \varepsilon_n^{ph}-\mu}\\
\times\psi_n(\x)\bar\psi_n(\x')\frac{e^{-i\omega_m (\tau-\tau')}}{\hbar\beta},
\end{multline}
\begin{multline}
\int_0^{\hbar\beta}d\tau''\int d{\bf x''}\:G_{ph(0)}(\x,\tau;\x'' ,\tau'')\\
\times G_{ph(0)}^{-1}(\x'',\tau'';\x ',\tau')
=\delta(\x-\x')\delta(\tau-\tau').
\end{multline}

\section{Coupled Gross-Pitaevskii-like equations}
We now turn to the Bose condensed exciton-photon system. Let us introduce the time independent order parameters $\chi_0(\x)$ and $\psi_0(\x)$ as
\be\label{chipsi3}
\chi(\x,\tau)=\chi_0(\x)+\chi'(\x,\tau),\quad
\psi(\x,\tau)=\psi_0(\x)+\psi'(\x,\tau),
\ee
where $\chi'(\x,\tau)$ and $\psi'(\x,\tau)$ are the fluctuations of the exciton and photon quantum fields around the order parameters $\chi_0(\x)$ and $\psi_0(\x)$. We shall use the Bogoliubov approximation assuming $\chi'(\x,\tau)\ll\chi_0(\x)$ $\psi'(\x,\tau)\ll\psi_0(\x)$, i.e. we confine ourselves to the case of low temperatures.

Substituting the definitions (\ref{chipsi3}) into the action given by (\ref{S}) and rearranging the derived terms according to their orders in $\chi'(\x,\tau)$ and $\psi'(\x,\tau)$, we obtain the following results.

The zeroth-order part of the action is given by  
\begin{multline}\label{FP}
\hbar\beta F_P[\chi_0,\bar\chi_0;\psi_0,\bar\psi_0]=
\hbar\beta\int\dx\left\{\frac{\hbar^2}{2m_{ex}}|{\bf \bigtriangledown}\chi_0(\x)|^2\right.\\
+\left(V^{ex}(\x)+E_0-\mu\right)|\chi_0(\x)|^2
+\frac{V_0}{2}|\chi_0(\x)|^4\\
+\frac{\hbar^2}{2m_{ph}}|{\bf \bigtriangledown}\psi_0(\x)|^2-\mu|\phi_0(\x)|^2\\
\left.+\frac{\hbar\Omega}{2}\left[\bar\psi_0(\x)\chi_0(\x)+\psi_0(\x)\bar\chi_0(\x)\right]\right\}
\end{multline}
and has the form of the Pitaevskii functional.

According to the semiclassical method, the functional reaches its minimum at functions $\chi_0(\x)$ and $\psi_0(\x)$, for which the functional variation vanishes. Equivalently, we can set the part of the action linear in $\chi'(\x,\tau)$ and $\psi'(\x,\tau)$ equal to zero. As a result, we obtain the set of coupled stationary equations analogous to the Gross-Pitaevskii equation:
\begin{multline}\label{GP}
\left(-\frac{\hbar^2\bigtriangledown^2}{2m_{ex}}+V^{ex}(\x)+E_0-\mu\right.\\
\left.\phantom{aaaa}+ V_0|\chi_0(\x)|^2\biggr)\chi_0(\x)
+\frac{\hbar\Omega}{2}\psi_0(\x)=0,\right.\\
\left(-\frac{\hbar^2\bigtriangledown^2}{2m_{ph}}-\mu\right)\psi_0(\x)+\frac{\hbar\Omega}{2}\chi_0(\x)=0.
\end{multline}
These equations provide a minimum of the functional (\ref{FP}) and describe the system of the two coupled condensates in the mean field approximation.

Let us discuss the derived equations by considering first the homogeneous case $V^{ex}(\x)=0$. In the absence of the Rabi splitting $\hbar\Omega=0$, the chemical potential of excitons equals $\mu=E_0+V_0 n_0^{ex}$, same as for the conventional one-component Bose condensed gas. Here we introduced the exciton condensate density $|\chi_0|^2=n_0^{ex}$. It is known that the Rabi splitting leads to the appearance of the two branches in the energy spectrum, which are referred to as the lower and upper polariton states (below we denote them by superscripts ``L'' and ``U'').  The chemical potential is equal to  
\be\label{mu}
\mu^{(L,U)}=\frac{1}{2}\left[(E_0+V_0\:n_0^{ex})\mp\sqrt{(E_0+V_0n_0^{ex})^2+\hbar^2\Omega^2}\right]
\ee
and varies within the interval $\mu^{(L)}\in[\mu_{min}^{(L)},0]$, $\mu^{(U)}\in[\mu_{min}^{(U)},\infty]$, where $\mu_{min}^{(L,U)}=(E_0\mp\sqrt{E_0^2+\hbar^2\Omega^2})/2$.  

In both cases the solution of the set of equations (\ref{GP}) has the form
\begin{align}\label{ChiPsi}
\chi_0&=\sqrt{\frac{1}{V_0}\frac{\mu^2-E_0\mu-\hbar^2\Omega^2/4}{\mu}},\notag\\
\psi_0&=\frac{\hbar\Omega}{2\mu}\sqrt{\frac{1}{V_0}\frac{\mu^2-E_0\mu-\hbar^2\Omega^2/4}{\mu}}.
\end{align}
For the lower polariton state, the exciton and photon condensate phases differ by $\pi$, since the chemical potential is always negative. For the upper polariton state, the exciton and photon condensate phases coincide and the chemical potential is positive.   

The energy of the homogeneous system per unit area, as it follows from (\ref{FP}), can be calculated as follows: 
\be\label{E}
E^{(L,U)}=E_0 n_{ex}+\frac{1}{2}V_0 (n_0^{ex})^2\mp\hbar\Omega\sqrt{n_0^{ex}n_0^{ph}},
\ee
where $n_0^{ph}=|\psi_0|^2$ is the photon condensate density.

Thus, when the homogeneous exciton-photon system is the condensate of lower polaritons, it has the minimal energy $E^{(L)}$. When the system is the condensate of upper polaritons, it has the maximal energy $E^{(U)}$. The difference between these energies per one particle is the energy of Rabi splitting $\hbar\Omega$ (in the limit of low density). The energies of the uniform states, for which phases of the exciton and photon condensates differ by a value ranging from $0$ to   $\pi$, are situated within the interval $[E^{(L)},E^{(U)}]$. Such states are time-dependent, since the exciton and photon condensates cyclically turn into each other, i.e. a kind of the internal Josephson effect takes place. We shall discuss these states elsewhere\cite{Voron3}. Here we confine ourselves only to the study of the lower polariton branch with the energy $E^{(L)}$, omitting further the superscript ``{\it L}'' for shortness. 

Substituting Eq. (\ref{ChiPsi}) into (\ref{E}), we get the connection between the energy and the chemical potential
\be\label{E2}
E=\frac{1}{2V_0}\left[\mu^2+\frac{1}{2}\hbar^2\Omega^2-E_0^2-\frac{\hbar^2\Omega^2}{\mu}E_0
-\frac{3}{16}\frac{\hbar^4\Omega^4}{\mu^2}\right].
\ee
We introduce the polariton condensate density as the sum
\be\label{nP}
n_0^P=n_0^{ex}+n_0^{ph},
\ee
Substituting Eq. (\ref{ChiPsi}) into (\ref{nP}), we find the connection between the polariton condensate density and the chemical potential 
\be\label{nP2}
n_0^P=\frac{1}{V_0}\left[\mu-E_0-\frac{1}{4}\frac{\hbar^2\Omega^2}{\mu^2}E_0-\frac{1}{16}\frac{\hbar^4\Omega^4}{\mu^3}\right].
\ee
With the help of Eqs. (\ref{E2}) and (\ref{nP2}) we can easily prove that $\partial E/\partial n_0^P=\mu$, i.e. the chemical potential is equal to the change in the energy as one extra polariton is added to the system. The fact that $\mu < 0$ means that the energy always decreases. It is worth pointing out that the expressions (\ref{ChiPsi}) could be obtained from the condition that the energy has a minimum at a fixed value of chemical potential $(\partial E/\partial n_0^{ex})_\mu=$ $(\partial E/\partial n_0^{ph})_\mu=0$. As it is seen from (\ref{ChiPsi}), in the limit of large  density $\mu\to 0$ the polariton condensate  becomes mainly photonic one, and the energy is now connected with the density by the relation $E=-(3/16)(\hbar^4\Omega^4/V_0)^{1/3}(n_0^P)^{2/3}$, which  differs principally from the relation for the conventional  one-component Bose condensate $E=(1/2)V_0 (n_0^{ex})^2$.   

Let us next discuss the inhomogeneous case. The lower equation in (\ref{GP}), describing the photon condensate, possesses the spatial scale which plays a role of the healing length and equals 
\be
\xi_{ph}=\frac{\hbar}{\sqrt{-2m_{ph}\mu}}.
\ee
The length $\xi_{ph}$ has a minimal value in the limit of zero density and increases with the chemical potential. 

For the upper equation in (\ref{GP}), describing the exciton condensate, the healing length equals
\be
\xi_{ex}=\frac{\hbar}{\sqrt{2m_{ex}(E_0-\mu)}}
\ee
and increases with the chemical potential as well.

It is important to point out that the relation 
$m_{ph}/m_{ex}\sim 10^{-4}$ observed typically in experiments causes the following inequality   
\be\label{xixi}
\xi_{ph}\gg \xi_{ex}.
\ee

Let's now study the conditions of the equivalence of the two-component exciton-photon and one-component polariton approaches. These approaches are obviously  equivalent in the homogeneous case, when $\psi_0=(\hbar\Omega/2\mu)\chi_0$, $n_0^P=|\psi_0|^2+|\chi_0|^2$. We expect these local relations to be valid also in the case of smooth densities change. We rewrite the lower equation in (\ref{GP}) in the integral form using the fact that the MacDonald function $K_0(x)$ (see, for example, \cite{Vladimirov})
is the Green's function of the equation. Thus we have  
\be\label{PsiChiCon}
\psi_0(\ra)=-\frac{1}{2\pi}\frac{\hbar\Omega}{2\mu}\int K_0(|\ra-\ra'|)\chi_0(\ra')d\ra'.
\ee
Here we introduced the dimensionless vector $\ra=\x/\xi_{ph}$. 
Taking the asympotics $K_0(x)\approx \sqrt{\pi/2x}e^{-x}$ valid for $x \to \infty$  into account, we can expand the smooth function $\chi(\ra')$ in a power series in the vicinity of $\ra'=\ra$ and keep only the first terms     
\be
\chi_0(\ra')\approx\chi(\ra)+\nabla\chi_0(\ra)\cdot\sa+\frac{1}{2} \sum_{i,j}\frac{\partial^2 \chi_0(\ra)}{\partial r_i\partial r_j }s_i s_j,
\ee
where $\sa=\ra'-\ra$. We substitute this expansion into (\ref{PsiChiCon}) and integrate it with respect to $d\sa=s\:ds\: d\varphi$. Here it is convenient to use the integral formula for the MacDonald function 
\be
K_0(x)=\int_1^\infty \frac{e^{-x\eta}}{\sqrt{\eta^2-1}}d\eta.
\ee
As a result, the gradient term and the term containing the second mixed partial derivative vanish after the integration with respect to the angle $d\varphi$, and we obtain 
\be
\psi_0(\x)=\frac{\hbar\Omega}{2\mu}\chi_0(\x)\left[1+\xi_{ph}^2\frac{\Delta\chi_0(\x)}{\chi_0(\x)}\right],
\ee
where $\Delta$ is the Laplace operator. Therefore the condition that determines the applicability of the local relations between particles densities is   
\be
\frac{\Delta\chi_0(\x)}{\chi_0(\x)}\ll \frac{1}{\xi_{ph}^2}
\ee
which means that the relative exciton density change must be small on the scales comparable to the photon healing length.

In topological structures such as vortices and solitons, the exciton density changes considerably on the scales of the exciton healing length. The same picture takes place near the boundaries of the condensate confined by the box potential.  In all these cases the one-component polariton approach is not applicable. 

The extensively used Thomas-Fermi approximation consists in neglecting the quantum pressure terms in the equations (\ref{GP})  at sufficiently high densities of condensate particles. The obtained estimate establishes the range of validity of Thomas-Fermi approximation. 

The solution of Eq.~(\ref{GP}) for the harmonic trap is published in Ref.\cite{Voron1}, the vortex solution is discussed in Ref.\cite{Voron2}. 

\section{Noncondensate particles}
\subsection{Matrix Green's function}
We can represent the quadratic part of the action $S^{quad}$ as a quadratic form
\begin{gather}
S^{quad}=-\frac{\hbar}{2}\Idtdx\;[\bar\chi',\bar\psi',\chi',\psi']\cdot{\bf G}^{-1}\cdot
\begin{bmatrix}\chi'\\ \psi'\\ \bar\chi'\\  \bar\psi'
\end{bmatrix},
\end{gather}
where the Green's function has a matrix structure
\begin{gather}
-{\bf G}=\Bigg\langle\begin{bmatrix}\chi'\\ \psi'\\ \bar\chi'\\ \bar\psi'
\end{bmatrix}\cdot[\bar\chi', \bar\psi',\chi',\psi']\Bigg\rangle
\end{gather}
and can be written as
\begin{multline}\label{G-1}
{\bf G}^{-1}(\x,\tau;\x',\tau')={\bf G}_{(0)}^{-1}(\x,\tau;\x',\tau')-\\
-\frac{1}{\hbar}
\begin{bmatrix}
2V_0|\chi_0(\x)|^2 & \hbar\Omega/2 & V_0\chi_0(\x)^2 & 0 \\
\hbar\Omega/2 & 0& 0 & 0 \\
V_0\bar\chi_0(\x)^2 & 0 &  2V_0|\chi_0(\x)|^2  & \hbar\Omega/2 \\
0 & 0 & \hbar\Omega/2 & 0
\end{bmatrix}\\
\times\delta(\x-\x')\delta(\tau-\tau'),
\end{multline}
where
\be
{\bf G}_{(0)}^{-1}=
\begin{bmatrix}
G_{ex(0)}^{-1}& 0 & 0 & 0 \\
0 & G_{ph(0)}^{-1}& 0 & 0 \\
0 & 0 & G_{ex(0)}^{-1}& 0 \\
0 & 0 & 0 & G_{ph(0)}^{-1}\\ 
\end{bmatrix}.
\ee
Here $G_{ex(0)}^{-1}(\x,\tau;\x',\tau')$ and $G_{ph(0)}^{-1}(\x,\tau;\x',\tau')$ are the inverse exciton and photon Green's functions given by (\ref{Gex0-1}) and (\ref{Gph0-1}), respectively.

We can rewrite (\ref{G-1}) as ${\bf G}^{-1}={\bf G}_{(0)}^{-1}-{\bf\Sigma}$, where ${\bf\Sigma}$ is a matrix formed by self-energies. Multiplying this equation from the left by ${\bf G}_{(0)}$ and from the right by ${\bf G}^{-1}$ we obtain the Dyson equation ${\bf G}={\bf G}_{(0)}+{\bf G}_{(0)}{\bf\Sigma}{\bf G}$. Some Green's functions from ${\bf G}$ are presented in Fig. 1. 
\begin{figure}[b]
\label{fig_Dyson_Equations}
\renewcommand{\captionlabeldelim}{.}
\includegraphics[width=0.9\columnwidth]{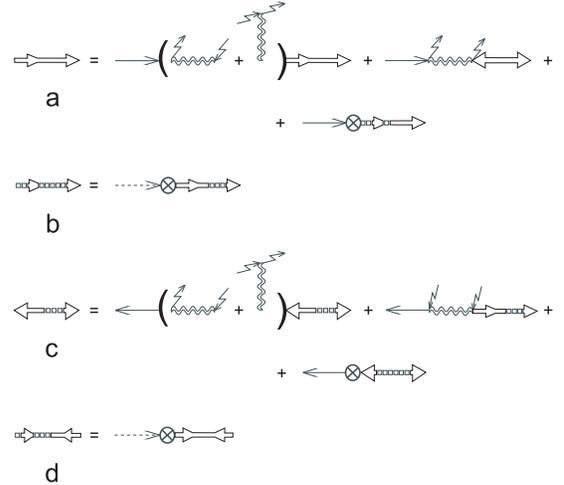}
\caption{\small  
Diagrammatic representation of some Dyson equations from ${\bf G}={\bf G}_{(0)}+{\bf G}_{(0)}{\bf\Sigma}{\bf G}$. Solid and dotted lines represent free-field exciton and photon Green's functions, respectively. Double wiggly line represents exciton-exciton interaction counted in the ladder approximation. Circle represents the Rabi splitting. (a) Exact exciton Green's function. (b) Exact photon Green's function. (c) Exact anomalous exciton-photon Green's function. (d) Exact anomalous photon-exciton Green's function. }
\end{figure}
We will only consider the case of a uniform system $V^{ex}(\x)=0$.  The exciton and photon single-particle energies have the form
\be
\varepsilon_n^{ex}=\varepsilon_\p^{ex}=\frac{\p^2}{2m_{ex}},\qquad\quad
\varepsilon_n^{ph}=\varepsilon_\p^{ph}=\frac{\p^2}{2m_{ph}}.
\ee
Substituting $\mu$, Eqs. (\ref{Gex0-1}) and (\ref{Gph0-1}) into (\ref{G-1}) gives
\begin{multline}\label{Gpk-1}
-\hbar{\bf G}^{-1}(\p,k)=\\
\begin{bmatrix}
E_\p^{ex}\hskip-4pt-i\hbar\omega_k & \hbar\Omega/2  & V_0\chi_0^2 & 0 \\
\hbar\Omega/2  & E_\p^{ph}\hskip-4pt-i\hbar\omega_k & 0 & 0 \\
V_0\bar\chi_0^2  & 0 & E_\p^{ex}\hskip-4pt+i\hbar\omega_k & \hbar\Omega/2  \\
0 & 0 & \hbar\Omega/2  & E_\p^{ph}\hskip-4pt+i\hbar\omega_k \\ 
\end{bmatrix},
\end{multline}
where we introduced the energies as
\begin{multline}\label{Eex}
E_\p^{ex}=\varepsilon_\p^{ex}+E_0-\mu+2V_0n_0=
\varepsilon_\p^{ex}+\frac{1}{2}\biggl(V_0 n_0\\
\left.-\sqrt{(E_0+V_0n_0)^2+\hbar^2\Omega^2}\right)+\left(\frac{E_0}{2}+V_0 n_0\right), 
\end{multline}
\begin{multline}\label{Eph}
E_\p^{ph}=\varepsilon_\p^{ph}-\mu=\varepsilon_\p^{ph}+\frac{1}{2}\left(V_0 n_0\right.\\
\left.-\sqrt{(E_0+V_0n_0)^2+\hbar^2\Omega^2}\right)-\left(\frac{E_0}{2}+V_0 n_0\right). 
\end{multline}
Hereafter, we denote $n_0^{ex}$ by  $n_0$ and replace $\chi_0$ by $\sqrt{n_0}\:e^{i\phi}$.

\subsection{Spectrum}
The determinant of the matrix ${\bf G}^{-1}(\p,\omega)$ is zero at the poles $\omega$ of the matrix ${\bf G}(\p,\omega)$. Equating the determinant to zero yields
\begin{multline}\label{detG}
\hbar^4\omega^4-\left({E_\p^{ex}}^2+{E_\p^{ph}}^2+\frac{\hbar^2\Omega^2}{2}-V_0^2n_0^2\right)\hbar^2\omega^2\\
+\left(\left(E_\p^{ex}E_\p^{ph}-\frac{\hbar^2\Omega^2}{4}\right)^2-V_0^2n_0^2{E_\p^{ph}}^2\right)=0.
\end{multline}
Solving Eq. (\ref{detG}) for $\omega^2$, we obtain the polariton energy spectrum modified by the condensate:
\begin{multline}\label{Spectrum}
\hbar^2\omega_\p^{(L,U)2}=\frac{1}{2}\left({E_\p^{ex}}^2+{E_\p^{ph}}^2-V_0^2n_0^2+\frac{\hbar^2\Omega^2}{2}\right)\\
\pm\frac{1}{2}\biggl[\left(-{E_\p^{ex}}^2+{E_\p^{ph}}^2+V_0^2n_0^2\right)^2\\
+\hbar^2\Omega^2\left(\left(E_\p^{ex}+E_\p^{ph}\right)^2-V_0^2n_0^2\right)\biggr].
\end{multline}
In the absence of the condensate $n_0=0$, it follows from Eqs. (\ref{Eex}), (\ref{Eph}) that $E_\p^{ex}=\varepsilon_\p^{ex}+E_0-\mu$, $E_\p^{ph}=\varepsilon_\p^{ph}-\mu$, and equation (\ref{Spectrum}) gives
\begin{multline}
\hbar^2\omega_\p^{(L,U)2}=\frac{1}{2}\biggl({(E_0+\varepsilon_\p^{ex}})^2+{\varepsilon_\p^{ph}}^2+\frac{\hbar^2\Omega^2}{2}+\mu^2\\
-2\mu(E_0+\varepsilon_\p^{ex}+\varepsilon_\p^{ph})\biggr)
\\
\pm\frac{1}{2}\left(E_0+\varepsilon_\p^{ex}+\varepsilon_\p^{ph}-2\mu\right)\\\times\sqrt{\left(E_0+\varepsilon_\p^{ex}-\varepsilon_\p^{ph}\right)^2+\hbar^2\Omega^2},
\end{multline}
which is equivalent to
\begin{multline}\label{LU}
\hbar\omega_\p^{(L,U)}=\frac{1}{2}\left(E_0+\varepsilon_\p^{ex}+\varepsilon_\p^{ph}\right)\\
\pm\frac{1}{2}\sqrt{\left(E_0+\varepsilon_\p^{ex}-\varepsilon_\p^{ph}\right)^2+\hbar^2\Omega^2}-\mu=\varepsilon_\p^{(L,U)}-\mu.   
\end{multline}                                                      
One can see that the spectrum turns into the energy dispersion of the lower and upper polaritons.

In the absence of the Rabi splitting we have
\begin{multline}
\hbar^2\omega_\p^2=\frac{1}{2}\left({E_\p^{ex}}^2+{E_\p^{ph}}^2-V_0^2n_0^2\right)\\
\pm\frac{1}{2}\left(-{E_\p^{ex}}^2+{E_\p^{ph}}^2+V_0^2n_0^2\right),
\end{multline} 
which gives the photon energy spectrum 
\be
\hbar\omega_\p=\pm E_\p^{ph}=\pm(\varepsilon_\p^{ph}-\mu)
\ee
and  the exciton energy spectrum modified by the condensate
\be\label{BogolSpectrum}
\hbar\omega_\p=\pm\sqrt{{E_\p^{ex}}^2-V_0^2n_0^2}=\pm\sqrt{{\varepsilon_\p^{ex}}^2+2V_0n_0\varepsilon_\p^{ex}}=\pm\varepsilon_\p^{B}. 
\ee
The equation (\ref{BogolSpectrum}) is the Bogoliubov dispersion law. Here we assume $\mu=E_0+V_0 n_0$ as discussed in Section 4.

For  small momenta $p\to 0$ the lower branch takes the phonon-like form
\be\label{phonon}
\hbar\omega_\p^{(L)}=v_s p,
\ee
where $v_s$ is the sound velocity. Assuming that $\alpha=m_{ph}/m_{ex}=0$, $V_0 n_0 \ll \hbar\Omega$, which is commonly encountered in practice, the velocity can be estimated as follows:
\be
v_s=\sqrt{\frac{\hbar^2\Omega^2}{E_0^2+\hbar^2\Omega^2}\frac{V_0 n_0}{4m_{ph}}}
\ee
At $p\to 0$ the upper branch  has the form
\be
\hbar\omega_\p^{(U)}=\Delta+\frac{p^2}{2m_0^{(U)}},
\ee
where  
\be
\Delta=\sqrt{E_0^2+\hbar^2\Omega^2}+\frac{1}{2}\left(1+\frac{3E_0}{\sqrt{E_0^2+\hbar^2\Omega^2}}\right)V_0 n_0
\ee
in the limit $V_0 n_0 \ll \hbar\Omega$, and
\begin{multline}
\frac{1}{m_0^{(U)}}=\Biggl[\frac{1}{2}\left(1-\frac{E_0}{\sqrt{E_0^2+\hbar^2\Omega^2}}\right)\\
-\frac{\hbar^2\Omega^2}{(E_0^2+\hbar^2\Omega^2)^{3/2}}V_0 n_0\Biggr]\frac{1}{m_{ph}}
\end{multline}
in the approximation $\alpha=m_{ph}/m_{ex}=0$, $V_0 n_0  \ll\hbar\Omega$.  

In the limit of large momenta $p\to \infty$ the lower polariton branch turns into the dispersion law of free excitons 
\begin{multline}
\hbar\omega_\p^{(L)}=\frac{p^2}{2m_{ex}}\\
+\frac{1}{2}\left(\sqrt{(E_0+V_0 n_0)^2+\hbar^2\Omega^2}+E_0+3V_0 n_0\right).
\end{multline}
If $\hbar\Omega=0$, the square root should be taken with the negative sign; this corresponds to $\mu=E_0+V_0 n_0$.

The upper branch for $p\to \infty$ takes the similar form
\begin{multline}
\hbar\omega_\p^{(U)}=\frac{p^2}{2m_{ph}}\\+\frac{1}{2}\left(\sqrt{(E_0+V_0 n_0)^2+\hbar^2\Omega^2}-(E_0+V_0 n_0)\right).
\end{multline}
The outlined properties of the polariton spectrum modified by the condensate are shown in Fig. 2. We see that the energy and momentum of the polariton resonance increase with the condensate density. This resembles a positive detuning between the exciton and photon modes. 
\begin{figure}[b]
\label{fig_Polariton_Spectrum}
\renewcommand{\captionlabeldelim}{.}
\includegraphics[width=1.0\columnwidth]{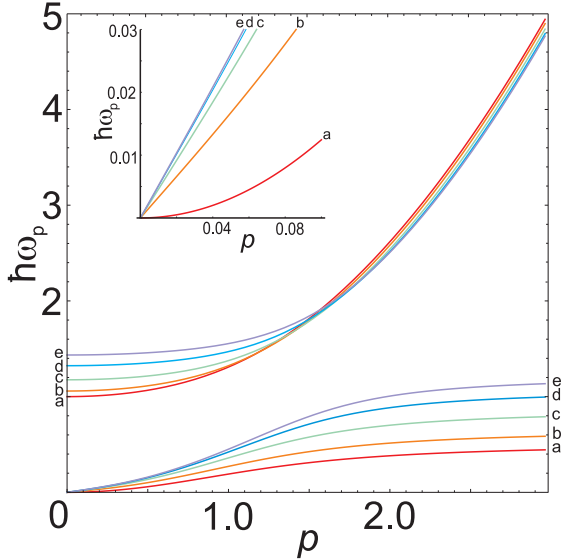}
\caption{\small  (Color online) Polariton energy spectrum modified by the condensate. The momentum is expressed in the units of $\sqrt{m_{ph}\hbar\Omega}$, the energy is expressed in the units of $\hbar\Omega$. (a)$V_0n_0=0$, (b)$V_0n_0=0.1\hbar\Omega$, (c)$V_0n_0=0.25\hbar\Omega$, (d)$V_0n_0=0.4\hbar\Omega$, (e)$V_0n_0=0.5\hbar\Omega$. Inset: phonon-like spectra in the vicinity of zero momentum. 
}
\end{figure}

\subsection{Polariton basis}
Let us now diagonalize the matrix ${\bf G}^{-1}$. The diagonalization does not reduce to straightforward determination of the eigenvalues because of opposite signs in front of $i\hbar\omega_k$ in the diagonal elements of the matrix (\ref{Gpk-1}). An option is to consider the generalized eigenvalue problem 
${\bf G}^{-1}\xi=(\hbar\omega-i\hbar\omega_k)\: {\bf W} \xi$, where ${\bf W}= diag(1,1,-1,-1)$ (see,  for example, {\cite{Golub}). However, we suggest a different approach: to multiply the two lower rows of ${\bf G}^{-1}$ by $-1$ and thus obtain the standard eigenvalue problem ${\bf \tilde G}^{-1}\xi=(\hbar\omega-i\hbar\omega_k)\: \xi$, where ${\bf \tilde G}^{-1}={\bf W}{\bf G}^{-1}$ is  a non-Hermitian operator  with respect to the scalar product $(a,b)=\sum_\alpha\bar a_\alpha b_\alpha$. Let us define the new scalar product as
\be\label{ab}
((a,b))=(a,{\bf W}b).
\ee
The Hermitian conjugation with respect to the old and new scalar products is given by the formulae 
\be
({\bf O}a, b)=(a,{\bf O}^+ b),\quad\quad (({\bf O}a, b))=((a,{\bf O}^{\bigoplus} b)).  
\ee
One can obtain the following relationship between the operators ${\bf O}^{\bigoplus}$ and ${\bf O}^+$:
\be
{\bf O}^{\bigoplus}={\bf W}{\bf O}^+{\bf W}.
\ee
The new scalar product is chosen in such a way that the equation $({\bf \tilde G}^{-1})^{\bigoplus}={\bf \tilde G}^{-1}$ holds. Therefore the operator ${\bf \tilde G}^{-1}$ is Hermitian with respect to the scalar product (\ref{ab}) and its eigenvalues are real: ${\bf \tilde G}^{-1}\xi=\lambda\xi$, $\lambda=\bar \lambda$ , if the norm of the vector $\xi$ is nonzero: $((\xi,\xi))\ne 0$. 

Let us perform the transformation from the exciton-photon basis $(\chi,\psi,\bar\chi,\bar\psi)$ to the polariton basis \linebreak $(P^{(L)},P^{(U)},\bar P^{(L)},\bar P^{(U)})$, where $P^{(L)}$ and $P^{(U)}$ describe the lower and upper polaritons, respectively. We shall carry out the canonical transformation  
\be
P_{\p}^{(L,U)}=\eta_{\chi_\p}^{(L,U)}\chi_\p+\eta_{\psi_\p}^{(L,U)}\psi_\p+\bar\nu_{\chi_\p}^{(L,U)}\bar\chi_\p+\bar\nu_{\psi_\p}^{(L,U)}\bar\psi_\p
\ee
where $\eta_{\chi_\p}^{(L,U)}$, $\eta_{\psi_\p}^{(L,U)}$, $\nu_{\chi_\p}^{(L,U)}$, and $\nu_{\psi_\p}^{(L,U)}$ are the unknown coefficients to be found. 

It is clear that 
\be
\bar P_{\p}^{(L,U)}=\bar\eta_{\chi_\p}^{(L,U)}\bar\chi_\p+\bar\eta_{\psi_\p}^{(L,U)}\bar\psi_\p+\nu_{\chi_\p}^{(L,U)}\chi_\p+\nu_{\psi_\p}^{(L,U)}\psi_\p
\ee
and therefore the two bases are related to each other by the equation 
\be
\begin{bmatrix}
P_{\p}^{(L)} \\ P_{\p}^{(U)} \\ \bar P_{\p}^{(L)} \\ \bar P_{\p}^{(U)}
\end{bmatrix}
=
\begin{bmatrix}
\eta_{\chi_\p}^{(L)} & \eta_{\psi_\p}^{(L)} & \bar\nu_{\chi_\p}^{(L)} & \bar\nu_{\psi_\p}^{(L)}\\   
\eta_{\chi_\p}^{(U)} & \eta_{\psi_\p}^{(U)} & \bar\nu_{\chi_\p}^{(U)} & \bar\nu_{\psi_\p}^{(U)}\\
\nu_{\chi_\p}^{(L)} & \nu_{\psi_\p}^{(L)} & \bar\eta_{\chi_\p}^{(L)} & \bar\eta_{\psi_\p}^{(L)}\\
\nu_{\chi_\p}^{(U)} & \nu_{\psi_\p}^{(U)} & \bar\eta_{\chi_\p}^{(U)} & \bar\eta_{\psi_\p}^{(U)}
\end{bmatrix}
\begin{bmatrix}
\chi_{\p} \\ \psi_{\p} \\ \bar\chi_{\p} \\ \bar\psi_{\p}
\end{bmatrix}.
\ee    
Let us denote the matrix in the right-hand side by B.

The diagonalization of a Hermitian matrix implies the use of a unitary matrix, which columns are normalized eigenvectors of the Hermitian matrix. Let us assume that the columns of the matrix ${\bf B}^+$ are the normalized eigenvectors of the operator ${\bf \tilde G}^{-1}$, i.e. we have  
\be\label{GB}
{\bf \tilde G}^{-1}{\bf B}^+={\bf B}^+{\bf \tilde G_d}^{-1},
\ee
where 
\begin{multline}
{\bf \tilde G_d}^{-1}=diag\left(\hbar\omega_\p^{(L)}-i\hbar\omega_k,\hbar\omega_\p^{(U)}-i\hbar\omega_k,\right.\\
\left.-\hbar\omega_\p^{(L)}-i\hbar\omega_k,-\hbar\omega_\p^{(U)}-i\hbar\omega_k\right).
\end{multline}
The columns of the matrix ${\bf B}^+$ are normalized with respect to the scalar product $((\bullet ))$, i.e. their components satisfy the equation 
\be\label{norm}
\left|\eta_{\chi_\p}^{(L,U)}\right|^2+\left|\eta_{\psi_\p}^{(L,U)}\right|^2-\left|\nu_{\chi_\p}^{(L,U)}\right|^2-\left|\nu_{\psi_\p}^{(L,U)}\right|^2=1.
\ee
The components of the eigenvectors can be found from the equations ${\bf \tilde G}^{-1}\xi=(\hbar\omega-i\hbar\omega_k)\: \xi$. We obtain the following relationships between the coefficients of the transformation:
\begin{align}\label{coef}
\eta_{\psi_\p}^{(L,U)}&=-\frac{\hbar\Omega/2}{E_\p^{ph}-\hbar\omega_\p^{(L,U)}}\:\eta_{\chi_\p}^{(L,U)},\notag\\
\nu_{\chi_\p}^{(L,U)}&=\frac{E_\p^{ph}+\hbar\omega_\p^{(L,U)}}{\hbar^2\Omega^2/4-(E_\p^{ex}+\hbar\omega_\p^{(L,U)})(E_\p^{ph}+\hbar\omega_\p^{(L,U)})},\notag\\
&\times V_0n_0e^{-2i\phi}\:\eta_{\chi_\p}^{(L,U)},\notag\\
\nu_{\psi_\p}^{(L,U)}&=-\frac{\hbar\Omega/2}{\hbar^2\Omega^2/4-(E_\p^{ex}+\hbar\omega_\p^{(L,U)})(E_\p^{ph}+\hbar\omega_\p^{(L,U)})},\notag\\
&\times V_0n_0e^{-2i\phi}\:\eta_{\chi_\p}^{(L,U)}.
\end{align}
Using (\ref{norm}) and (\ref{coef}) we can find 
$\eta_{\chi_\p}^{(L,U)}$, $\eta_{\psi_\p}^{(L,U)}$, $\nu_{\chi_\p}^{(L,U)}$, and $\nu_{\psi_\p}^{(L,U)}$.

We present another method for the calculation of the transformation's coefficients.
From Eq. (\ref{GB}) we obtain
\begin{multline}
{\bf \tilde G}^{-1}={\bf B}^+{\bf \tilde G_d}^{-1}({\bf B}^+)^{-1}={\bf B}^+{\bf \tilde G_d}^{-1}({\bf B}^+)^{\bigoplus}\\
={\bf B}^+{\bf \tilde G_d}^{-1}{\bf W}{\bf B}{\bf W}.
\end{multline}
If the matrix ${\bf  G}^{-1}={\bf W}{\bf \tilde G}^{-1}$ corresponded to the diagonal matrix ${\bf  G_d}^{-1}={\bf W}{\bf \tilde G_d}^{-1}$, we would write
\be
{\bf  G}^{-1}={\bf B}^{-1}{\bf  G_d}^{-1}{\bf W}{\bf B}{\bf W}.
\ee 
From this equation, it follows that 
\be\label{GG}
{\bf  G}={\bf B}^+{\bf  G_d}{\bf B},
\ee
where
\begin{multline}
{\bf  G_d}=diag\left(\frac{1}{-i\hbar\omega_k+\hbar\omega_\p^{(L)}},\frac{1}{-i\hbar\omega_k+\hbar\omega_\p^{(U)}},\right.\\
\left.\frac{1}{i\hbar\omega_k+\hbar\omega_\p^{(L)}},\frac{1}{i\hbar\omega_k+\hbar\omega_\p^{(U)}}\right).
\end{multline}
Multiplying the matrices ${\bf  G_d}$, ${\bf B}$ and ${\bf B}^+$ according to (\ref{GG}), we express the components of the matrix ${\bf  G}$ through the components of ${\bf B}$. Thus, for example, 
\begin{multline}\label{G11}
G_{11}(\p,k)=\frac{\left|\eta_{\chi_\p}^{(L)}\right|^2}{-i\hbar\omega_k+\hbar\omega_\p^{(L)}}+\frac{\left|\nu_{\chi_\p}^{(L)}\right|^2}{i\hbar\omega_k+\hbar\omega_\p^{(L)}}\\+\frac{\left|\eta_{\chi_\p}^{(U)}\right|^2}{-i\hbar\omega_k+\hbar\omega_\p^{(U)}}+\frac{\left|\nu_{\chi_\p}^{(U)}\right|^2}{i\hbar\omega_k+\hbar\omega_\p^{(U)}}, 
\end{multline}
\begin{multline}\label{G22}
G_{22}(\p,k)=\frac{\left|\eta_{\psi_\p}^{(L)}\right|^2}{-i\hbar\omega_k+\hbar\omega_\p^{(L)}}+\frac{\left|\nu_{\psi_\p}^{(L)}\right|^2}{i\hbar\omega_k+\hbar\omega_\p^{(L)}}\\+\frac{\left|\eta_{\psi_\p}^{(U)}\right|^2}{-i\hbar\omega_k+\hbar\omega_\p^{(U)}}+\frac{\left|\nu_{\psi_\p}^{(U)}\right|^2}{i\hbar\omega_k+\hbar\omega_\p^{(U)}}. 
\end{multline}
On the other hand, we can find the matrix ${\bf  G}$ directly by inverting the matrix  ${\bf  G}^{-1}$ from Eq. (\ref{Gpk-1}). Thus, for 
$G_{11}$ and $G_{22}$ we have
\begin{multline}
G_{11}(\p,k)=-\frac{\hbar}{D}\left[E_\p^{ex}\hbar^2\omega_k^2+\left({E_\p^{ph}}^2+\hbar^2\omega_k^2\right.\right.\\
\left.\left.+\frac{(\hbar\Omega)^2}{4}\right)i\hbar\omega_k
+E_\p^{ex}{E_\p^{ph}}^2-E_\p^{ph}\frac{(\hbar\Omega)^2}{4}\right],
\end{multline}
\begin{multline}
G_{22}(\p,k)=\\
-\frac{\hbar}{D}\left[E_\p^{ph}\hbar^2\omega_k^2+\left({E_\p^{ex}}^2+\hbar^2\omega_k^2\right.\right.\\
\left.+\frac{(\hbar\Omega)^2}{4}-V_0^2n_0^2\right)i\hbar\omega_k+{E_\p^{ex}}^2E_\p^{ph}\\
\left.-E_\p^{ex}\frac{(\hbar\Omega)^2}{4}-E_\p^{ph}V_0^2n_0^2\right],
\end{multline}
where
\begin{multline}
D=(i\hbar\omega_k-\hbar\omega_\p^L)(i\hbar\omega_k-\hbar\omega_\p^U)\\ \times(i\hbar\omega_k+\hbar\omega_\p^L)(i\hbar\omega_k+\hbar\omega_\p^U).
\end{multline}

Bringing these equations to the form similar to that of Eqs. (\ref{G11}) and (\ref{G22}), we obtain the coefficients $\eta_{\chi_\p}^{(L,U)}$, $\eta_{\psi_\p}^{(L,U)}$, $\nu_{\chi_\p}^{(L,U)}$, and $\nu_{\psi_\p}^{(L,U)}$.  

Both methods lead to the following expressions for the coefficients of the transformation from the basis of excitons and photons to the basis of lower and upper polaritons in the presence of the condensate:
\begin{widetext} 
\begin{multline}\label{PLU}
P_\p^{(L,U)}=e^{i\xi^{(L,U)}}\frac{1}{\sqrt{2\hbar\omega_\p^{(L,U)}}}\frac{1}{\sqrt{\hbar^2\omega_\p^{(U)2}-\hbar^2\omega_\p^{(L)2}}}\times\\
\times\left\{\sqrt{\frac{\hbar^2\Omega^2/4-\left(E_\p^{ex}+\hbar\omega_\p^{(L,U)}\right)\left(E_\p^{ph}+\hbar\omega_\p^{(L,U)}\right)}{\hbar\omega_\p^{(L,U)}-E_\p^{ph}}}\left[\left(\hbar\omega_\p^{(L,U)}-E_\p^{ph}\right)\chi_\p+\frac{\hbar\Omega}{2}\psi_\p\right]+\right.\\
\left.+V_0 n_0\: e^{2i\phi}\sqrt{\frac{\hbar\omega_\p^{(L,U)}-E_\p^{ph}}{\hbar^2\Omega^2/4-\left(E_\p^{ex}+\hbar\omega_\p^{(L,U)}\right)\left(E_\p^{ph}+\hbar\omega_\p^{(L,U)}\right)}}\left[\left(\hbar\omega_\p^{(L,U)}+E_\p^{ph}\right)\bar\chi_\p-\frac{\hbar\Omega}{2}\bar\psi_\p\right]\right\}.
\end{multline}
\end{widetext}
Here $\xi^{(L,U)}$ are the phases which up to an arbitrary phase $\xi$ satisfy the equation $\xi^{(L)}=\xi^{(U)}-\pi/2=\xi$.

In a sense, these transformations incorporate both the Hopfield transformation for polaritons and the Bogoliubov transformation for weakly-interacting Bose gas.

For $\hbar\Omega=0$ the expression (\ref{PLU}) in the case of lower polariton yields  
\begin{multline}\label{bogoltrans}
P_\p^{(L)}=-e^{i\xi}\sqrt{\frac{\varepsilon_\p^{ex}+V_0n_0+\varepsilon_\p^B}{2\varepsilon_\p^B}}\chi_\p\\
+e^{2i\phi+i\xi}\sqrt{\frac{\varepsilon_\p^{ex}+V_0n_0-\varepsilon_\p^B}{2\varepsilon_\p^B}}\bar\chi_\p.
\end{multline}
The coefficients of $\chi_\p$ и $\bar\chi_\p$ are the coefficients $u_\p$ and $v_\p$ of the Bogoliubov transformation. 
An upper polariton turns into a photon: $P_\p^{(U)}=e^{i\xi}\psi_\p$. For $V_0 n_0=0$ the expression (\ref{PLU}) yields  
\begin{multline}
P_\p^{(L,U)}=\frac{\varepsilon_\p^{(L,U)}-\varepsilon_\p^{ph}}{\sqrt{\left(\varepsilon_\p^{ph}-\varepsilon_\p^{(L,U)}\right)^2+\hbar^2\Omega^2/4}}\chi_\p\\
+\frac{\hbar\Omega/2}{\sqrt{\left(\varepsilon_\p^{ph}-\varepsilon_\p^{(L,U)}\right)^2+\hbar^2\Omega^2/4}}\psi_\p.
\end{multline}
Here the phase $\xi$ is assumed to be zero. The coefficients of $\chi_\p$ and $\psi_\p$ are the Hopfield coefficients.

In Fig. 3 we plot the momentum dependences of the calculated coefficients. It is seen that the coefficients for the lower polariton similar to the Bogoliubov transformation coefficients have a singularity at zero momentum. This is due to the fact that, by virtue of Eq. (\ref{phonon}), $\hbar\omega_\p^{(L)}$ in the denominator of Eq. (\ref{PLU}) behaves like  $v_sp$.   In addition, the presence of the condensate shifts the balance between the excitons and photons to the photons in both polariton branches. Apparently, it is a consequence of the positive detuning between the exciton and photon modes discussed in Section 5.2. It is worth pointing out that exciton-photon interaction causes an extremely slow decay of the coefficient $\nu_{\chi_\p}^{(L)}$.
\begin{figure}[b]
\label{fig_Polariton_Coefficients}
\renewcommand{\captionlabeldelim}{.}
\includegraphics[width=\columnwidth]{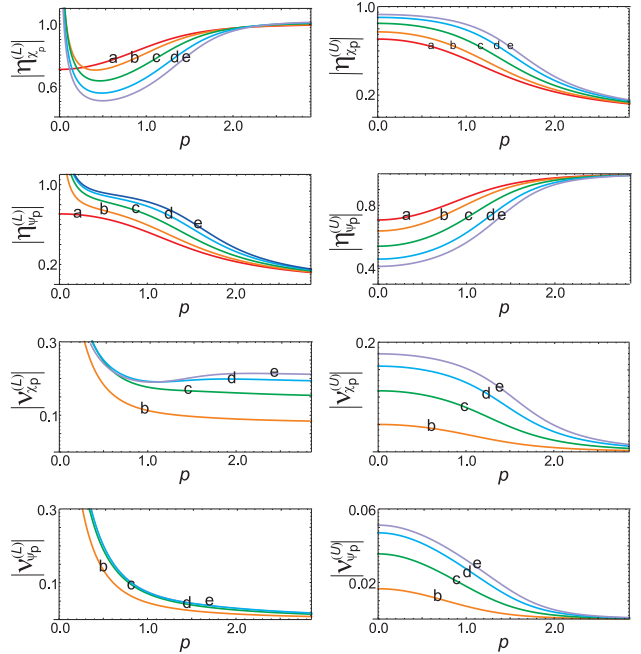}
\caption{\small (Color online) The transformation coefficients from excitons and photons to lower and upper polaritons as functions of momentum. The momentum is expressed in the units of $\sqrt{m_{ph}\hbar\Omega}$. (a)$V_0n_0=0$, (b)$V_0n_0=0.1\hbar\Omega$, (c)$V_0n_0=0.25\hbar\Omega$, (d)$V_0n_0=0.4\hbar\Omega$, (e)$V_0n_0=0.5\hbar\Omega$.}
\end{figure}

\subsection{Occupation numbers}
We can find the noncondensate exciton density taking into account that $n^{ex'}=- G_{11}(\x,\tau,\x',\tau^{+})$. Using Eq. (\ref{G11}), we obtain
\begin{multline}
n^{ex'}=-\sum_{\p\ne0,k}G_{11}(\p,k)=\\
=-\sum_{\p\ne0,k}\left\{\frac{\left|\eta_{\chi_\p}^{(L)}\right|^2}{-i\hbar\omega_k+\hbar\omega_\p^{(L)}}+\frac{\left|\nu_{\chi_\p}^{(L)}\right|^2}{i\hbar\omega_k+\hbar\omega_\p^{(L)}}\right.\\
\left.+\frac{\left|\eta_{\chi_\p}^{(U)}\right|^2}{-i\hbar\omega_k+\hbar\omega_\p^{(U)}}+\frac{\left|\nu_{\chi_\p}^{(U)}\right|^2}{i\hbar\omega_k+\hbar\omega_\p^{(U)}}\right\}. 
\end{multline}
To evaluate the sum over Matsubara frequencies, we use the formula 
\be\label{matsubarasum}
\lim_{\eta\to 0}\frac{1}{\hbar\beta}\sum_{k}\frac{e^{i\omega_k\eta}}{i\omega_k-(\varepsilon-\mu)/\hbar}=-\frac{1}{e^{\beta(\varepsilon-\mu)}-1}
\ee
and find that  
\begin{multline}\label{nex}
n^{ex'}=\sum_{\p\ne0}\left\{\frac{\left|\eta_{\chi_\p}^{(L)}\right|^2+\left|\nu_{\chi_\p}^{(L)}\right|^2}{e^{\beta\omega_\p^{(L)}}-1}+\left|\nu_{\chi_\p}^{(L)}\right|^2
\right\}\\
+\sum_{\p\ne0}\left\{\frac{\left|\eta_{\chi_\p}^{(U)}\right|^2+\left|\nu_{\chi_\p}^{(U)}\right|^2}{e^{\beta\omega_\p^{(U)}}-1}+\left|\nu_{\chi_\p}^{(U)}\right|^2
\right\}
\end{multline}
Substituting the explicit expressions for the coefficients $\eta_{\chi_\p}^{(L,U)}$, $\eta_{\psi_\p}^{(L,U)}$, $\nu_{\chi_\p}^{(L,U)}$, $\nu_{\psi_\p}^{(L,U)}$, we can write this equation as follows:
\be
n^{ex'}=\frac{1}{V}\left(\sum_{\p\ne 0}n_\p^{exL'}+\sum_{\p\ne 0}n_\p^{exU'}\right),
\ee
where the first term contains the sum over exciton occupation numbers of the lower polariton branch
\begin{multline}\label{nexL}
n_\p^{exL'}=-\frac{{-E_\p^{ph}}^2+\hbar^2{\omega_\p^L}^2-(\hbar\Omega)^2/4}{2\left(\hbar^2{\omega_\p^L}^2-\hbar^2{\omega_\p^U}^2\right)}\\
+\frac{-E_\p^{ex}{E_\p^{ph}}^2+E_\p^{ex}\hbar^2{\omega_\p^L}^2+E_\p^{ph}(\hbar\Omega)^2/4}{\hbar^2{\omega_\p^L}^2-\hbar^2{\omega_\p^U}^2}\\
\times\frac{1}{\hbar\omega_\p^L}\left(\frac{1}{2}+\frac{1}{e^{\beta\hbar\omega_\p^L}-1}\right)
\end{multline}
and the second sum is over exciton occupation numbers of the upper polariton branch
\begin{multline}\label{nexU}
n_\p^{exU'}=\frac{{-E_\p^{ph}}^2+\hbar^2{\omega_\p^U}^2-(\hbar\Omega)^2/4}{2\left(\hbar^2{\omega_\p^L}^2-\hbar^2{\omega_\p^U}^2\right)}\\
-\frac{-E_\p^{ex}{E_\p^{ph}}^2+E_\p^{ex}\hbar^2{\omega_\p^U}^2+E_\p^{ph}(\hbar\Omega)^2/4}{\hbar^2{\omega_\p^L}^2-\hbar^2{\omega_\p^U}^2}\\
\times\frac{1}{\hbar\omega_\p^U}\left(\frac{1}{2}+\frac{1}{e^{\beta\hbar\omega_\p^U}-1}\right).
\end{multline}
Similarly, for photons one has 
\begin{multline}\label{nph}
n^{ph'}=- G_{22}(\x,\tau,\x',\tau^+)=-\sum_{\p\ne0,k}G_{22}(\p,k)=\\
=-\sum_{\p\ne0,k}\left\{\frac{\left|\eta_{\psi_\p}^{(L)}\right|^2}{-i\hbar\omega_k+\hbar\omega_\p^{(L)}}+\frac{\left|\nu_{\psi_\p}^{(L)}\right|^2}{i\hbar\omega_k+\hbar\omega_\p^{(L)}}\right.\\
\left.+\frac{\left|\eta_{\psi_\p}^{(U)}\right|^2}{-i\hbar\omega_k+\hbar\omega_\p^{(U)}}+\frac{\left|\nu_{\psi_\p}^{(U)}\right|^2}{i\hbar\omega_k+\hbar\omega_\p^{(U)}}\right\}=\\
=\sum_{\p\ne0}\left\{\frac{\left|\eta_{\psi_\p}^{(L)}\right|^2+\left|\nu_{\psi_\p}^{(L)}\right|^2}{e^{\beta\omega_\p^{(L)}}-1}+\left|\nu_{\psi_\p}^{(L)}\right|^2
\right\}\\
+\sum_{\p\ne0}\left\{\frac{\left|\eta_{\psi_\p}^{(U)}\right|^2+\left|\nu_{\psi_\p}^{(U)}\right|^2}{e^{\beta\omega_\p^{(U)}}-1}+\left|\nu_{\psi_\p}^{(U)}\right|^2
\right\} 
\end{multline}
which leads to
\be
n^{ph'}=\frac{1}{V}\left(\sum_{\p\ne 0}n_\p^{phL'}+\sum_{\p\ne 0}n_\p^{phU'}\right),
\ee
where
\begin{multline}\label{nphL}
n_\p^{phL'}=-\frac{{-E_\p^{ex}}^2+\hbar^2{\omega_\p^L}^2-(\hbar\Omega)^2/4+V_0^2n_0^2}{2\left(\hbar^2{\omega_\p^L}^2-\hbar^2{\omega_\p^U}^2\right)}\\
+\frac{-{E_\p^{ex}}^2E_\p^{ph}+E_\p^{ph}\hbar^2{\omega_\p^L}^2+E_\p^{ex}(\hbar\Omega)^2/4+E_\p^{ph}V_0^2n_0^2}{\hbar^2{\omega_\p^L}^2-\hbar^2{\omega_\p^U}^2}\\
\times\frac{1}{\hbar\omega_\p^L}\left(\frac{1}{2}+\frac{1}{e^{\beta\hbar\omega_\p^L}-1}\right),
\end{multline}
\begin{multline}\label{nphU}
n_\p^{phU'}=\frac{{-E_\p^{ex}}^2+\hbar^2{\omega_\p^U}^2-(\hbar\Omega)^2/4+V_0^2n_0^2}{2\left(\hbar^2{\omega_\p^L}^2-\hbar^2{\omega_\p^U}^2\right)}\\
-\frac{-{E_\p^{ex}}^2E_\p^{ph}+E_\p^{ph}\hbar^2{\omega_\p^U}^2+E_\p^{ex}(\hbar\Omega)^2/4+E_\p^{ph}V_0^2n_0^2}{\hbar^2{\omega_\p^L}^2-\hbar^2{\omega_\p^U}^2}\\
\times\frac{1}{\hbar\omega_\p^U}\left(\frac{1}{2}+\frac{1}{e^{\beta\hbar\omega_\p^U}-1}\right).
\end{multline}
Here $n_\p^{phL'}$ and $n_\p^{phU'}$ are the photon occupation numbers of the lower and upper polariton branches.

Let us now analyze the results. In the absence of the Rabi splitting the first term in Eq. (\ref{nexL}) equals $1/2$ and the brackets' coefficient in the second term turns into $(\varepsilon_\p^{ex}+V_0n_0)/(\hbar\omega_\p)$, where $\hbar\omega_\p$ is given by Eq. (\ref{BogolSpectrum}). Both terms in Eq. (\ref{nexU}) are equal to zero, i.e. there are no excitons in the upper branch. As a result, for excitons we obtain the expression
\begin{multline}
n_\p^{ex'}=n_\p^{exL'}+n_\p^{exU'}=\\
=\frac{\varepsilon_\p+V_0n_0-\hbar\omega_\p}{2\hbar\omega_\p}+\frac{\varepsilon_\p+V_0n_0}{\hbar\omega_\p}\frac{1}{e^{\beta\hbar\omega_\p}-1}
\end{multline}
which is the standard momentum distribution of noncondensate particles in the Bogoliubov model.
 
Both terms in (\ref{nphL}) are equal to zero, i.e. there are no photons in the lower polariton branch. The first term in (\ref{nphU}) equals $-1/2$ and the brackets' coefficient equals $1$. Thus for photons we obtain  
\be
n_\p^{ph'}=n_\p^{phL'}+n_\p^{phU'}=\frac{1}{e^{\beta\hbar\omega_\p}-1}
\ee
which is the standard distribution of particles in the ideal Bose gas.
\begin{figure}[b]
\label{fig_Noncondensate_Density}
\renewcommand{\captionlabeldelim}{.}
\includegraphics[width=\columnwidth]{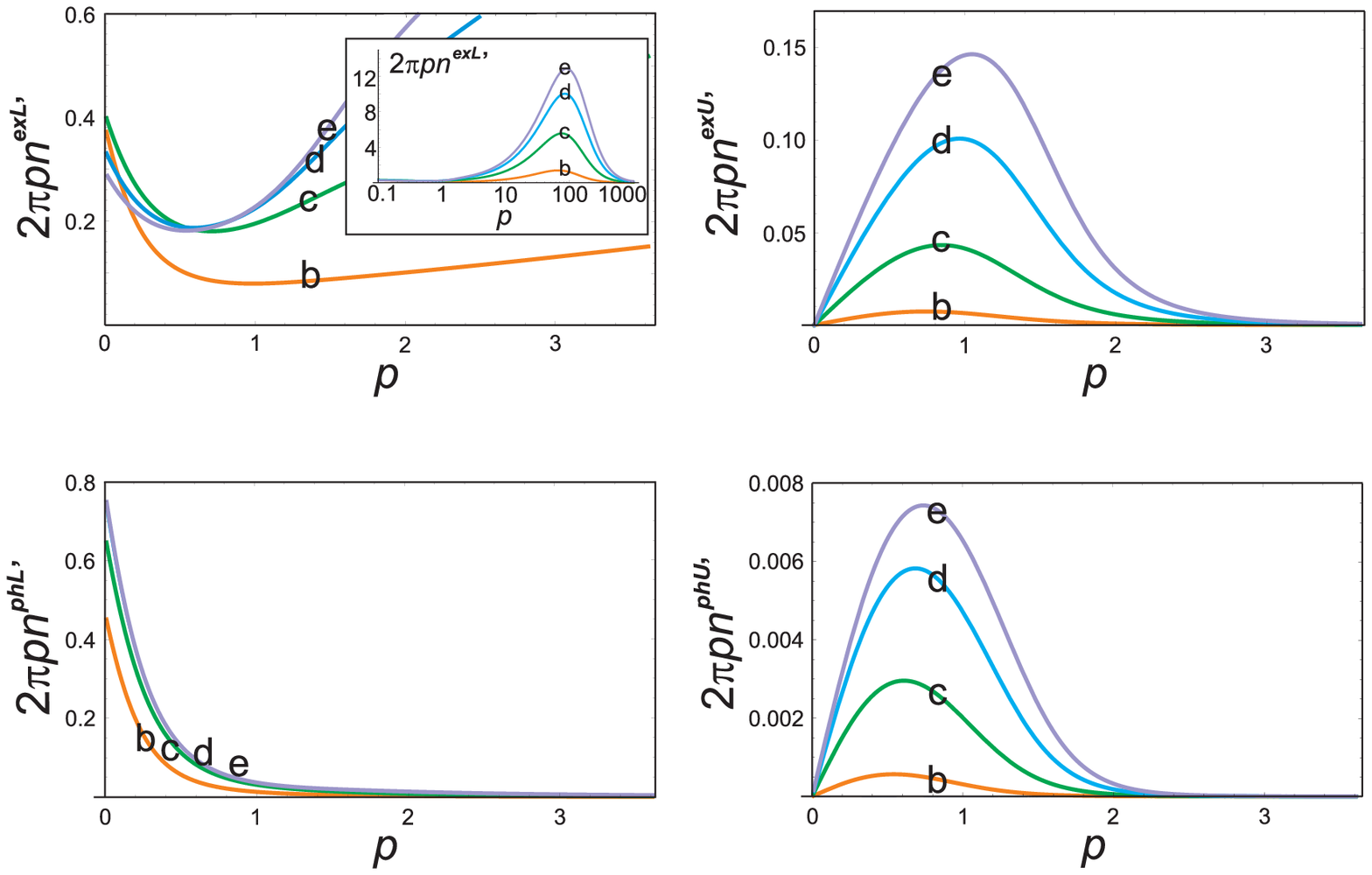}
\caption{\small (Color online) Exciton and photon momentum distributions in the lower and upper polariton branches. Momentum is expressed in the units of $\sqrt{m_{ph}\hbar\Omega}$. (a)$V_0n_0=0$, (b)$V_0n_0=0.1\hbar\Omega$, (c)$V_0n_0=0.25\hbar\Omega$, (d)$V_0n_0=0.4\hbar\Omega$, (e)$V_0n_0=0.5\hbar\Omega$. The inset shows the exciton momentum distribution in the lower polariton branch at large momenta (of the order of $\sqrt{m_{ex}/m_{ph}}\sim 10^2$).
}
\end{figure}
In Fig. 4 we plot the noncondensate exciton and photon distributions at zero temperature. At small momenta the exciton distribution is linearly decreasing starting from some finite value:
\begin{multline}
2\pi p n_\p^{exL'}\approx \frac{\pi}{4}\frac{V_0 n_0}{v_s}\left(1-\frac{E_0}{\Delta}\right)^2\\
-\frac{\pi}{2}\left(1-\frac{E_0}{\Delta}\right)\left[1-\left(1-\frac{E_0}{\Delta}\right)\frac{V_0n_0}{\Delta}\right]p, 
\end{multline}
but then it starts increasing with growth of momentum. Such behavior is a consequence of very slow drop of the coefficient $\nu_{\chi_\p}^{(L)}$ (see the previous subsection). The further behavior of the exciton noncondensate distribution is shown in the inset of Fig. 4,a. It is seen that the distribution drops off at very large momenta.  It is worth noting that the population of the upper polariton branch is high (comparable with that of the low polariton branch). This population increases with the condensate density.  The maximum of the population corresponds to the region of the polariton resonance   and shifts to large momenta with the growth of the condensate density.

\section{Coupled Bogoliubov-de-Gennes-like equations}
In order to investigate the inhomogeneous case we introduce the operators
\be
\hat K_{ex}=-\frac{\hbar^2\bigtriangledown^2}{2m_{ex}}+V^{ex}(\x)-\mu,
\ee
\be
\hat K_{ph}=-\frac{\hbar^2\bigtriangledown^2}{2m_{ph}}-\mu
\ee
and solve the following eigenvalue problem:
\begin{multline}\label{BdGmatrix}
\begin{bmatrix}
\hat K_{ex}+2V_0|\chi_0(\x)|^2 & \hbar\Omega/2 & V_0(\chi_0(\x))^2 & 0 \\
\hbar\Omega/2 & \hat K_{ph} & 0 & 0 \\
V_0(\bar\chi_0(\x))^2 & 0 &  \hat K_{ex}+2V_0|\chi_0(\x)|^2  & \hbar\Omega/2 \\
0 & 0 & \hbar\Omega/2 & \hat K_{ph}
\end{bmatrix}\\
\times\begin{bmatrix} u_n^{ex}(\x)\\ u_n^{ph}(\x) \\ v_n^{ex}(\x) \\ v_n^{ph}(\x)
\end{bmatrix}=\hbar\omega_n\begin{bmatrix}
1 & 0 & 0 & 0 \\
0 & 1 & 0 & 0 \\
0 & 0 &  -1  & 0 \\
0 & 0 & 0 & -1
\end{bmatrix}
\begin{bmatrix} u_n^{ex}(\x)\\ u_n^{ph}(\x) \\ v_n^{ex}(\x) \\ v_n^{ph}(\x)
\end{bmatrix}.
\end{multline}
Thus, we obtain a set of equations which is an analog of the Bogoliubov -- de Gennes system of equations and describes excitations in the system of coupled condensates of photons and excitons:
\begin{align}\label{BdG}
\left(-\frac{\hbar^2\bigtriangledown^2}{2m_{ex}}+V^{ex}(\x)-\mu+2V_0n_0(\x)\right)&u_n^{ex}(\x)\notag\\
+V_0(\chi_0(\x))^2v_n^{ex}(\x)+\frac{\hbar\Omega}{2}u_n^{ph}(\x)&=\hbar\omega_n u_n^{ex}(\x),\notag\\
\left(-\frac{\hbar^2\bigtriangledown^2}{2m_{ph}}-\mu\right)u_n^{ph}(\x)+\frac{\hbar\Omega}{2}u_n^{ex}(\x)&=\hbar\omega_n u_n^{ph}(\x),\notag\\ 
\left(-\frac{\hbar^2\bigtriangledown^2}{2m_{ex}}+V^{ex}(\x)-\mu+2V_0n_0(\x)\right)&v_n^{ex}(\x)\notag\\
+V_0(\bar\chi_0(\x))^2u_n^{ex}(\x)+\frac{\hbar\Omega}{2}v_n^{ph}(\x)&=-\hbar\omega_n v_n^{ex}(\x),\notag\\
\left(-\frac{\hbar^2\bigtriangledown^2}{2m_{ph}}-\mu\right)v_n^{ph}(\x)+\frac{\hbar\Omega}{2}v_n^{ex}(\x)&=-\hbar\omega_n v_n^{ph}(\x).
\end{align}

Similarly to Eq. (\ref{nex}), one can write an expression for  noncondensate excitons
\be\label{nex2}
n^{ex'}(\x)=\sum_{n\ne0}\left\{\frac{\left|u_n^{ex}(\x)\right|^2+\left|v_n^{ex}(\x) \right|^2}{e^{\beta\omega_n}-1}+\left|v_n^{ex}(\x)\right|^2
\right\}. 
\ee
and for photons
\be\label{nph2}
n^{ph'}(\x)=\sum_{n\ne0}\left\{\frac{\left|u_n^{ph}(\x)\right|^2+\left|v_n^{ph}(\x) \right|^2}{e^{\beta\omega_n}-1}+\left|v_n^{ph}(\x)\right|^2
\right\}. 
\ee
Functions $u_n^{ex}(\x)$, $v_n^{ex}(\x)$, $u_n^{ph}(\x)$ and 
$v_n^{ph}(\x)$ are normalized as
\begin{multline}
\int d\x\left(\left|u_n^{ex}(\x)\right|^2+\left|u_n^{ph}(\x)\right|^2-\right.\\
\left.\left|v_n^{ex}(\x) \right|^2-\left|v_n^{ph}(\x) \right|^2\right)=1.
\end{multline}
This equation is analogous to Eq. (\ref{norm}) and can be obtained in the same manner.

The solutions of the set of equations (\ref{BdG}) are out of the scope of the present discussion and will be published elsewhere.

\section{Conclusions}
In conclusion, the central result of the work is the canonical transformation from the basis of excitons and photons to the basis of lower and upper polaritons in the presence of the condensate. The transformation coefficients we obtained can be called the Hopfield coefficients renormalized by the condensate.

The path integral method we used has essentially more abilities than were demonstrated in the work. The method opens an effective way to analyze the formation and dynamics of a non-equilibrium condensate. The two-component approach based on the path integral method leads to the two-component diagram technique, which was briefly presented (Fig. 1), but not applied. All this aspects will be discussed in future works. 

\section{Acknowledgments}
Authors are grateful to Nina Voronova for help with preparation of this manuscript.

\end{document}